# Efficient self-imaging grating couplers on a Lithium-Niobate-On-Insulator platform at near-visible and telecom wavelengths

Emma Lomonte[1,2,3], Francesco Lenzini[1,2,3*], and Wolfram H.P. Pernice[1,2,3*]

[1] *Institute of Physics, University of Münster, Wilhelm-Klemm-Straße 10, 48149 Münster, Germany*
[2] *CeNTech – Center for Nanotechnology, Heisenbergstraße 11, 48149 Münster, Germany*
[3] *SoN – Center for Soft Nanoscience, Busso-Peus-Straße 10, 48149 Münster, Germany*

\* lenzini@uni-muenster.de, wolfram.pernice@uni-muenster.de

**Lithium-Niobate-On-Insulator (LNOI) has emerged as a promising platform in the field of integrated photonics. Nonlinear optical processes and fast electro-optic modulation have been reported with outstanding performance in ultra-low loss waveguides. In order to harness the advantages offered by the LNOI technology, suitable fiber-to-chip interconnects operating at different wavelength ranges are demanded. Here we present easily manufacturable, self-imaging apodized grating couplers, featuring a coupling efficiency of the $TE_0$ mode as high as $\simeq$47.1% at λ=1550 nm and $\simeq$44.9% at λ=775 nm. Our approach avoids the use of any metal back-reflector for an improved directivity or multi-layer structures for an enhanced grating strength.**

## INTRODUCTION

For applications that require a $\chi^{(2)}$ dielectric response, Lithium Niobate ($LiNbO_3$, LN) has always represented one of the materials of choice because of its remarkable bulk properties, such as a wide transparency range (350<λ<5200 nm), and large electro-optic ($r_{33}$=33 pm/V) and nonlinear optical ($d_{33}$=27 pm/V) coefficients. In the last decades, LN waveguides have been predominantly fabricated by ion-exchange processes[1], but suffered from a low refractive index contrast (Δn<0.1) between core and cladding, which results in weak modal confinement and bulky optical circuits with large (>1 mm) bending radii. As an attractive alternative, single-crystal LN thin films bonded onto a $SiO_2$ insulation layer (LNOI: Lithium-Niobate-On-Insulator) have been recently developed via the smart-cut technique[2], and allow for the realization of high index contrast (Δn$\simeq$0.7) waveguides by using electron beam lithography and dry etching methods. LNOI waveguides provide small footprint (bending radii <100 μm) with propagation loss as low as $\simeq$2.7 dB/m and $\simeq$6 dB/m in the telecom and visible range, respectively[3,4]. Furthermore, the stronger confinement of light in the guiding material has allowed for the realization of electro-optic modulators operating at a speed up to $\simeq$100 GHz with a CMOS-compatible voltage[5], as well as for achieving an unprecedented level of conversion efficiency in second-order nonlinear optical processes[6,7].

Based on these properties, LNOI stands out as a very attractive material platform for the implementation of complex integrated photonic circuits[8], especially for applications in quantum photonic technologies, where a large number of components for the generation, manipulation, and detection of quantum states of light is required on-chip and, ideally, integrated in a single device[9–14]. To take full advantage of the excellent performance demonstrated on-chip, a key requirement is the development of efficient fiber-to-chip interfaces. These are needed to join LNOI circuits with optical communication networks and to realize low-loss interconnects between optical systems with different functionalities[15]. To optically access the photonic circuitry, edge or vertical couplers are the two most widely utilized strategies. Depending on the application and the additional constraints – such as wideband operation, polarization sensitivity, or pliant positioning of the couplers on the photonic chip – one of the two approaches is favourable over the other[16].

While conventional $LiNbO_3$ ion-exchanged waveguides embedded in the bulk crystal feature an optical beam size of around 10 μm, standard LNOI butt couplers have shown substantial insertion loss in the range of 4.5-10 dB[5,17,18], due to the large mode mismatch between the optical fiber and the waveguide. Although more refined structures, consisting of a bilayer tapered mode size converter, provide insertion loss of only 1.7 dB/facet[19], they still rely on the use of lensed optical fibers to achieve high coupling efficiency, and are therefore sensible to micron-scale misalignments between the fiber core and the waveguide facet. A further improvement to this strategy has been recently demonstrated by use of a silicon oxynitride cladding to cover the LN end-fire couplers. This approach allows to better match the mode field distribution of the employed ultra-high numerical aperture optical fiber, reaching an optimal coupling efficiency of approximately -0.55 dB for both polarizations[20]. Nevertheless, edge couplers lack the possibility of coupling light in the direction normal to the waveguide plane, which is preferred for large-scale prototyping of densely integrated photonic circuits.

Diffractive grating structures, whose periodicity redirects the light beam from the waveguide plane toward a fiber array placed atop the photonic chip, constitute the most common type of vertical couplers. Unlike silicon photonics, where optimized gratings with insertion loss <1 dB have been experimentally demonstrated[21–24], to date grating couplers implemented in the less mature LNOI technology show reduced performance[25–34]. At telecom wavelengths, apodized grating couplers implemented on a pure LNOI platform – i.e., without the use of any back-reflector to grant improved directivity[25,29,34] or additional material layers to increase the grating strength[27,28,32] – have exhibited a maximum coupling efficiency of -3.6 dB at $\lambda \simeq 1550$ nm[33]. The use of a specially fabricated LNOI wafer with a 100 nm thick Au back-reflector has recently enabled to boost the efficiency of these gratings, and to achieve a coupling efficiency of 72% and 61.6% for TE- and TM-polarized light, respectively [34]. At visible and near-visible wavelengths only a shallow investigation has been conducted, and grating couplers showed insertion loss as high as 5.7 dB at $\lambda \simeq 780$ nm[11].

We note that an alternative strategy for the realization of efficient and broadband vertical couplers can be provided by polymeric structures fabricated by 3D direct-laser writing[35,36]. However, in addition to a more sophisticated fabrication procedure, these couplers can only be integrated with ridge waveguides. This prevents their usage for several applications of LNOI circuits, where a rib geometry – i.e., a partially etched waveguide – is often needed to achieve phase-matching in nonlinear optical processes[37] or to realize high-speed electro-optic modulators[17].

In this work we present our theoretical and experimental results on self-imaging grating couplers in the LNOI platform, operating in the near-visible (n-VIS) and in the infrared (IR) wavelength ranges. We demonstrate an alternative and flexible strategy for the fabrication of efficient grating couplers that can be easily adapted to match any desired film thickness and waveguide etching depth. Our grating structures feature a focusing effect of the upward-diffracted beam at a vertical distance of approximately 100-150 μm from the circuitry plane, made possible by use of a negative diffraction angle rather than the more conventional positive one[38–40]. This imaging effect allows to spatially match the mode supported by a single-mode optical fiber even with structures characterized by a low grating strength, which require the use of long gratings for achieving a high diffraction efficiency (see the Section on Theoretical background and device simulation for further details).

To highlight the flexibility of our approach, we realize grating couplers operating at both 775 nm and 1550 nm wavelengths, which are concurrently manufactured with a single electron beam lithographic exposure and etching step. This approach is suitable for the implementation of telecom-wavelength parametric-down conversion sources in LNOI waveguides[13], where efficient couplers for both the pump beam and the signal and idler photons are required on the same chip. By use of a 300 nm thick X-cut LN film bonded on a Silica-on-Silicon wafer, we experimentally demonstrate a coupling efficiency of the fundamental TE mode as high as -3.48 dB at $\lambda \simeq 775$ nm and -3.27 dB at $\lambda \simeq 1550$ nm. Furthermore, we numerically estimate that by complementing our gratings with a metal back-reflector, our approach can allow for the achievement of a near unitary coupling efficiency at both wavelengths.

**THEORETICAL BACKGROUND AND DEVICE SIMULATION**

Diffractive grating couplers (GCs) allow for phase-matching between the light confined into the plane of the photonic circuit and the mode at the output of an optical fiber, via a periodic modulation of the refractive index of the guiding material in one or more directions, i.e., along the direction of light propagation (1D-GCs) or also along the width of the waveguide (2D-GCs)[41]. As we are interested in waveguides with a single polarization state, we consider only 1D-GCs, which can effectively couple light in the fundamental $TE_0$ mode in and out of the LN waveguide. Indeed, due to the LN birefringence, 1D-GCs are naturally polarization-selective and do not allow to simultaneously couple orthogonal polarization states at the same central wavelength and with the same diffraction angle. With reference to Fig. 1a-b, 1D-GCs obey the Bragg equation[16]:

$$n_{air} \sin \theta = n_{eff} - m \frac{\lambda}{\Lambda}, \text{ with } n_{eff} = FF \cdot n_{LN\ tooth} + n_{trench} \cdot (1 - FF),$$

where $\Lambda$ is the grating period, m the diffraction order, $\lambda$ the wavelength of the light in vacuum, $n_{air}$ the refractive index of the material in which the gaussian mode exiting the fiber is propagating through, and $n_{eff}$ is the effective refractive index of the light in the grating structure. As a first-order approximation, $n_{eff}$ can be expressed as a weighted average of the effective index of the light in the unetched LN tooth, named as $n_{LN\ tooth}$, and in the etched trench, named as $n_{trench}$. The weight is given by the filling factor FF, defined as FF=d/$\Lambda$. Here, $\theta$ represents the diffraction angle of the optical beam. In the input configuration, this is the angle at which the optical beam leaving the optical fiber impinges on the grating coupler. By assuming $\theta=0°$ for the direction normal to the optical circuitry, we call the angle positive or negative for a clockwise or a counterclockwise rotation with respect to the normal direction.

In uniform GCs – namely gratings characterized by a constant $\Lambda$ and FF over the whole periodic structure – most of the light is emitted at the beginning of the grating, resulting in an exponential decay of the upward-emitted light and poor overlap with the Gaussian mode of an optical fiber, substantially clipping the achievable overall coupling efficiency. Apodization – meaning a variation of the filling factor along the propagation length of the grating – can be exploited to locally tailor the grating strength, thus obtaining a diffracted beam with a Gaussian-like shape[42]. Nonetheless, a drawback of photonic platforms such as LNOI or $Si_3N_4$ is that, due to their relatively low index contrast and grating strength, a diffraction efficiency close to unity requires the use of long gratings with overall dimensions that are much larger than the mode field diameter (MFD) of a single-mode optical fiber. To circumvent this fundamental limitation and to realize efficient grating couplers integrated with LNOI waveguides, we adopt a strategy already demonstrated for $Si_3N_4$ waveguides[38–40], which allows for the creation of a self-focusing effect of the diffracted Gaussian beam at a vertical distance in the range of 50-100 μm above the chip. This approach paves the way for the realization of GCs where a high diffraction efficiency and an almost unitary overlap with the mode of an optical fiber can be obtained simultaneously. Importantly, this self-focusing effect can only be achieved along both the transverse and the longitudinal direction of the grating (see the Section on Device fabrication for further details) by use of grating couplers designated for phase-matching the first diffraction order with a negative diffraction angle.

In Fig. 1a-b we present a sketch of the structure employed for the simulation and fabrication of our gratings. Our material stack consists of a rib waveguide with an etching depth of 255 nm, which is patterned along the Y-axis of a 300 nm thick X-cut LN thin film, bonded onto a 4.75 μm thick $SiO_2$ insulation layer thermally grown on a Si substrate. Due to the physical etching process utilized to fabricate our LN circuit, the waveguides and the grating teeth show a typical sidewall angle of $\simeq 62°$. A hydrogen silsesquioxane (HSQ) layer is used as an upper cladding (see the Section on Device fabrication for further details). The same film thickness and etching depth are used to simulate and realize grating couplers operating at a central wavelength of $\lambda \simeq 775$ nm and of $\lambda \simeq 1550$ nm.

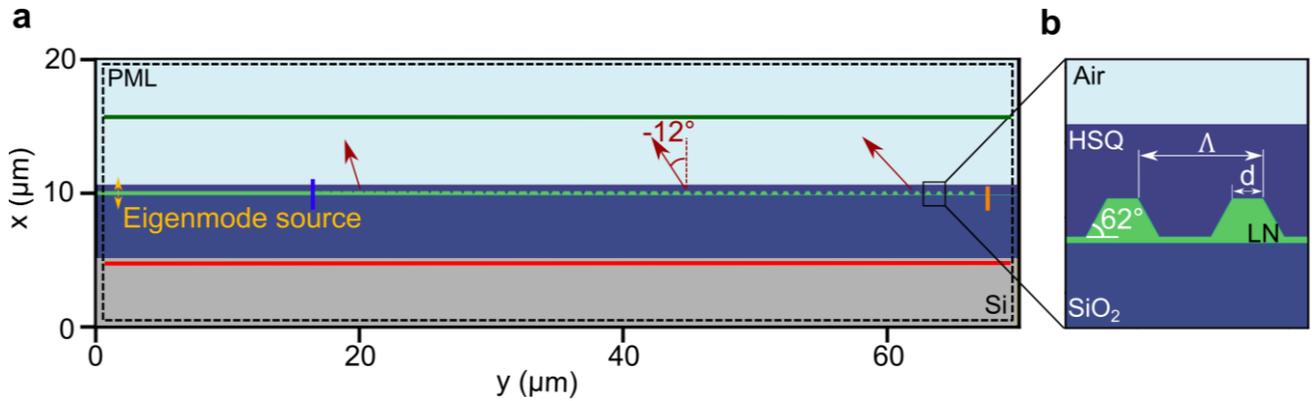

**Fig. 1. Sketch of the structure employed for the simulation and fabrication of the grating couplers. (a)** The material stack consists of HSQ-cladded (purple) LN optical circuitry (green) on top of a $SiO_2$ layer (blue) grown on a Si substrate (grey). The dashed line around the cell corresponds to the perfectly matched layer (PML) used to implement the absorbing boundary condition. The eigenmode source, illustrated as a yellow double-ended arrow, is placed inside the LN waveguide, and propagates through the grating coupler, where the apodization on the filling factor (FF) locally tailors the diffraction angle (red arrows) such that at the center of the grating the upward-diffracted light is emitted with a -12° angle to match the polishing angle of the employed fiber array. Depicted are also the monitor planes used to compute the Poynting flux spectra of the upward- and downward-diffracted electromagnetic radiation (green and red line, respectively), the fraction of light that leaks into the LN slab at the end of the grating (orange line), and the light back-reflected into the waveguide (blue line). **(b)** Zoom on two grating teeth, where the FF is defined as the ratio between the top width of the grating tooth d and the grating period Λ. Due to the physical sputtering process used to etch the photonic circuits, the grating teeth are modeled with a sidewall angle of 62°.

To find the best geometrical parameters of our etched GCs, we run 2D finite domain time difference (FDTD) simulations on Meep, a free and open-source software for electromagnetic simulations[43]. Without loss of generality, we restrict our theoretical study to the outcoupling scheme, where a $TE_0$ light source is located inside the waveguide and propagates through the scattering structure where it gets diffracted (see Fig. 1a). Indeed, grating couplers can be modeled as linear open systems, for which the reciprocity theorem is valid, implying an identical coupling efficiency for both the input and the output configuration[16]. The simulation cell is discretized with a homogeneous grid, whose resolution is set to a value for which at least three elementary cells correspond to the finest structure. A 1 μm thick perfectly matched layer (PML) is inserted all around the cell to implement absorbing boundary conditions (dashed line in Fig. 1a). PMLs absorb with no reflection all the light impinging on them, regardless of the frequency and the angle of incidence[44], hence allowing to simulate an infinite space with a finite simulation cell.

To evaluate the efficiency of our couplers, we perform two different kinds of simulation. First, by using a continuous-wave eigenmode source, we compute the optical power $|E|^2$ and optimize the grating period and the filling factor, such that the light can be diffracted upward with an angle of approximately -12°, to match the 8° polishing angle of our in-house fiber array. For simplicity, we keep the grating period constant over the whole length of the grating coupler, and only vary the FF. To realize a self-focusing effect for the upward diffracted light we take advantage of the fact that, by varying the FF along the propagation length, we do not only locally change the grating strength, but also the diffraction angle (see Fig. 1a and Fig. 2a,d). The initial FF is set equal to the largest value feasible with our fabrication process, while minimizing the difference in the effective index of the light at the interface between the waveguide and the grating, and then we decrease it linearly to a value for which the desired MFD is obtained in correspondence of the beam waist. We introduce ten additional uniform grating periods after the apodization section to reduce the index mismatch with the surrounding area at the end of the grating coupler region.

To estimate the overall coupling efficiency, besides the MFD of the diffracted mode, we also calculate the grating directivity, the grating strength, and the amount of back-reflected light at the input of the grating. With reference to Fig. 1a, these parameters can be estimated by launching a Gaussian eigenmode source and computing the Poynting flux spectra of the upward- and downward-diffracted electromagnetic radiation (green and red monitor, respectively), the fraction of light that leaks into the LN slab at the end of the grating (orange monitor), and the light back-reflected into the waveguide (blue monitor). We iterate this comprehensive analysis until a proper combination of all the GC parameters is found to maximize the upward-diffracted light and to simultaneously minimize the back-reflection, which is caused by the index mismatch between the grating structure and the waveguide and results in parasitic Fabry-Pérot oscillations[16].

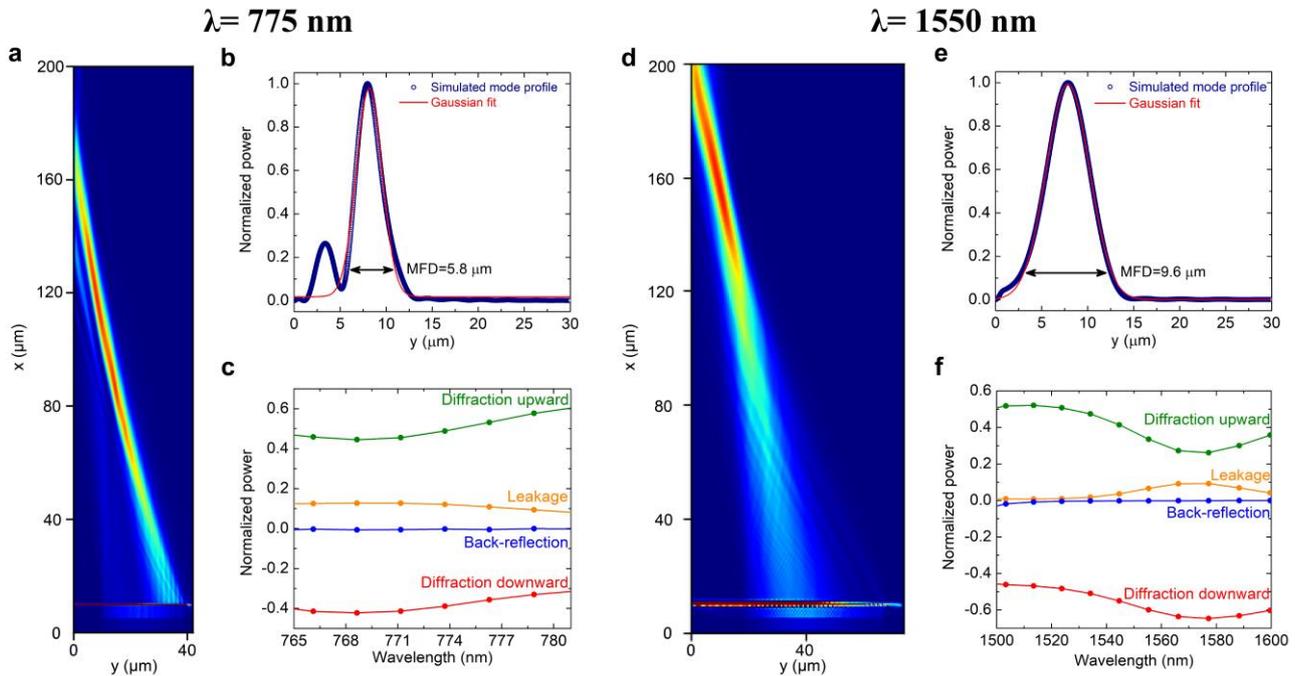

**Fig. 2. Results of our FDTD simulations in the near-visible (left panel) and infrared (right panel) wavelength ranges. (a-d)** Optical power $|\mathbf{E}|^2$ computed in the outcoupling scheme. **(b-e)** Simulated mode field profile at the beam waist (blue curve), fitted with a Gaussian function (red line). Reported are also the estimated MFDs. **(c-f)** Calculated Poynting flux spectra from the monitor planes of Fig. 1a. The sign of the calculated flux spectra is determined by the direction of the normal vector of the monitor planes.

In Fig. 2, we show the results of our simulation. In both the n-VIS and the IR wavelength ranges, the diffracted beam is emitted at an angle of approximately -12°, with a clear focusing effect on the longitudinal direction at a height of ≃110 and ≃150 μm from the waveguide plane, respectively (see Fig. 2a-d). At 775 nm wavelength, this is achieved with a grating period of 370 nm and an apodization on the FF from an initial value equal to 75% down to 25% across seventy teeth. At λ=1550 nm, we use fifty grating teeth over which we keep the grating period constant at 830 nm, while linearly diminishing the FF from 85% at the beginning of the grating to 25% at the end. In both cases, this apodization correctly results in a Gaussian-like form of the upward-emitted beam, as shown in Fig. 2b-e. By fitting the simulated mode profile at its beam waist with a Gaussian function, we derived a MFD equal to ≃5.8 μm at λ=775 nm, and to ≃9.6 μm at λ=1550 nm. These values lead to an overlap integral >99% with the optical mode supported by the 780HP single-mode optical fiber (MFD≃5.5 μm at λ=775 nm), and the SMF28 single-mode fiber (MFD≃10.4 μm at λ=1550 nm), respectively. In the case of the grating coupler for near-visible light, we attribute the presence of the secondary peak in the upward-diffracted beam to the destructive interference between the light diffracted at the coupler/SiO$_2$ interface and the light back-reflected at the SiO$_2$/Si interface. Indeed, this side-lobe is hardly visible when running the simulation without the Si substrate (data not showed).

In Fig. 2c-f we report the simulated Poynting flux spectra of our grating couplers. At telecom wavelengths they have been computed in the range 1500-1600 nm; in the near-visible spectrum, the wavelength range over which they have been computed is restricted to the one accessible with the tunable laser source employed in our experiment (see the Section on Experimental results for further details). The oscillation of the flux spectra in dependence of the wavelength is due to the interference between the light diffracted at the grating surface and the light reflected at the $SiO_2$/Si interface. Although the given $SiO_2$ thickness does not provide constructive interference at the two central wavelengths of 1550 nm and 775 nm, a high coupling efficiency of approximately 50% is obtained at around 1510 nm in the infrared regime, while in the near-visible range, a CE as high as 60% can be estimated around 780 nm. Unlike the case of near-visible wavelengths, where near-zero back-reflections have been computed over the entire wavelength range, it can be observed that in the infrared regime this contribution becomes slightly more definite at shorter wavelengths.

It is well known that an almost perfect grating directivity can be achieved by adding a metal back-reflector between the Si handle and the $SiO_2$ insulation layer, which allows for successfully redirecting the light diffracted down into the substrate in the upper direction[23,24]. To corroborate the advantages of our coupling strategy, as a proof-of-principle we computed the coupling efficiency of the presented GCs by using the parameters described above and a 100 nm thick Al film. We estimate that more than 95% of light impinging on the grating is diffracted upward, while maintaining the desired Gaussian-like profile (results not reported). We choose Aluminum due to its reflectivity >90% over the entire visible and infrared light spectrum; additionally, because of its good adhesion to the glass layer, Al back-reflectors have been already experimentally demonstrated by back wet-etching of the Si substrate[23].

## RESULTS AND DISCUSSION

**Device fabrication.** The starting point of our fabrication process is a diced LNOI 15x15 mm² chip consisting of a 300 nm thick X-cut LN film bonded onto a 4.75 μm thick buried $SiO_2$ insulation layer, thermally grown on a Si handle (from NanoLN). To fabricate the photonic circuitry and the in-out standard-tooth couplers, a single electron exposure and etching step are performed. Negative-tone resist (ArN7520) is patterned by means of a 100 keV e-beam lithography system. Subsequently, we develop the resist in a MF-319 solution, and a low pressure Ar plasma is used to transfer the pattern into the LN thin film in an Oxford100 ICP-RIE etching tool by a physical sputtering process[3]. We choose an etching depth equal to $\simeq$255 nm, leaving a slab of $\simeq$45 nm behind. Albeit the purely physical sputtering is effective in providing smooth etched waveguide surfaces, it comes with two main aftereffects: the fabricated waveguides show a typical sloped profile with a sidewall angle of approximately 62° (measured by atomic force microscopy); and the etched material can redeposit on the waveguide surfaces. To remove the sidewall redeposition, after stripping the resist, we immerse the photonic chip in an RCA-1 cleaning bath[45]. Further, to minimize the scattering loss, we clad the optical circuitry with a 750 nm thick HSQ layer – a well-known flowable oxide that can transform into glass when properly thermally or electrically cured[46].

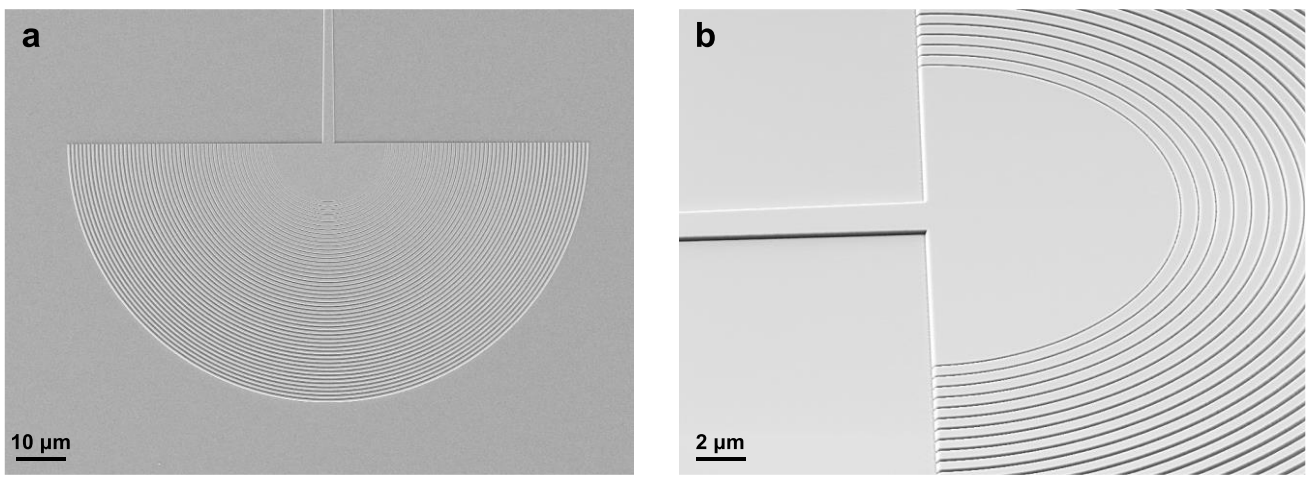

**Fig. 3. Scanning electron micrographs of a grating coupler operating at IR wavelengths. (a)** View from the top. **(b)** Closeup image of the grating coupler taken in proximity of the tapered region, characterized by an opening angle of 180°.

In Fig. 3 we show two scanning-electron microscopy (SEM) micrographs of a grating coupler operating in the infrared wavelength range, taken before covering the LN circuitry with the HSQ upper cladding. To realize an imaging effect also in the transverse direction, our gratings are implemented as a series of concentric rings with a radius equal to the distance from the input waveguide[38]. A taper with an opening angle of 180°, chosen to ensure that light is not confined along the transverse direction while propagating inside the coupler, is used to connect the grating to the waveguide circuit. With reference to Fig. 2a,d, we set the length of the taper equal to the distance between the first grating tooth and the horizontal position of the longitudinal focus of the diffracted beam calculated with our FDTD simulations. This way, we can ensure that the light diffracted by the grating is focused exactly into the same point along both the longitudinal and the transverse direction[38]. We design the top width of the waveguide at the input of the grating to obtain the desired MFD also in the transverse direction, in correspondence of the focal point of the diffracted beam. Since light is not laterally confined in our couplers and propagates in free-space after being diffracted by the grating, we can calculate the required waveguide width by using standard analytical formulas for the propagation of Gaussian beams in free-space and dielectric media. By using a finite-element mode solver for calculating the fundamental TE mode supported by our waveguides, we determined that this is achieved with a top waveguide width equal to 2.5 μm at $\lambda$=1550 nm and to 1.5 μm at $\lambda$=775 nm.

**Experimental results.** To characterize the performance of our devices, the fabricated photonic circuit consists of two identical grating couplers connected by a short waveguide. To simultaneously ensure transverse focus at $\lambda$=775 nm and single-mode (SM) operation of our waveguides, the top waveguide width at the intersection with the grating coupler is set equal to 1.5 μm and is then decreased down to 200 nm with an 80 μm long taper, whose length is chosen to guarantee adiabatic transition without the excitation of higher order modes. Analogously, at $\lambda$=1550 nm the waveguide starts with a top width of 2.5 μm that is reduced to 660 nm over a taper length of 90 μm. The distance between the two gratings is set equal to 127 μm, to match the pitch of the employed V-groove fiber array, which comprises of eight SM optical fibers at 1550 nm and eight SM channels at 775 nm. The fiber array is mounted on a goniometer stage that allows us to change the angle of the optical fibers to match the one of the diffracted light beams emitted by the GCs.

Continuous-wave (CW) laser light coming from a tunable laser source (either a TLB-6712 Newport Velocity diode laser with 765-781 nm tuning range or a TSL-550 Santec laser with 1500-1630 nm tuning range) propagates through a 3-paddle fiber polarization controller used to ensure TE input polarization and then it is coupled to one of the two grating couplers. The light leaving the photonic chip through the second coupler is collected via a Newfocus fiber-coupled photoreceiver and the wavelength response is monitored by continuously sweeping the laser. To assess the coupling efficiency per grating coupler, we normalize the recorded transmission spectrum by measuring the laser output as a function of the wavelength. Additionally, we subtract from the measured transmission the insertion loss of the employed fiber array, and the propagation loss of the light in the waveguide. Hence, assuming an identical behavior of the two grating couplers, the coupling efficiency is given by the square root of the properly normalized total transmission of the device. The results of our transmission measurements are reported in Fig. 4.

As an interesting feature of our couplers, optical beams with different central wavelength are diffracted at slightly different angles and thus can be coupled in the same device by shifting the position of the fiber array along the longitudinal direction of the grating. In the two panels of Fig. 4 we report several transmission spectra recorded from the same GC, which are obtained by optimizing the coupler transmission with a different alignment wavelength ($\lambda_{align}$). At visible wavelengths, due to the limited tuning range of our laser, for the same device we plot only two different transmission spectra. Here, although the highest coupling efficiency is determined to be $\simeq$44.9% for the spectrum recorded at the central wavelength $\lambda_{align}$=775 nm, no sensible difference could be observed in the case of $\lambda_{align}$=770 nm. The 3 dB bandwidth of the coupler is determined to be equal to approximately 11 nm.

At telecom wavelengths, a more complex behavior is observed. Analogously to the analysis performed in the near-visible wavelength range, here we show four different spectra for four different alignment wavelengths. As predicted by our simulations (see Fig. 2f), the maximum coupling efficiency (CE) is obtained around $\lambda$=1510 nm, reaching a CE as high as $\simeq$47.1%. Additionally, the stronger Fabry-Pérot oscillations for wavelengths below 1520 nm accurately resemble the simulated contribution of the back-reflected light, which is more dominant at shorter wavelengths. The 3 dB bandwidth of the coupler is determined to be equal to approximately 35 nm.

Even though the experimental measurements well agree with the simulated results, a slight reduction of the measured coupling efficiency compared to the theoretical values is found. In the telecom band, at $\lambda$=1510 nm we experimentally determined a coupling efficiency of 47.1% against the simulated value of 52%, given by the product of the normalized light upward-emitted by the GC and the bidimensional overlap integral between the Gaussian-like diffracted beam and the optical mode at the output of an optical fiber. Analogously, at $\lambda$=775 nm, we simulated a coupling efficiency of approximately 50%, while the efficiency determined by the measurement is $\simeq$44.9%. We attribute this small deviation to either a non-ideal transverse focus along the transverse direction, which was designed with a solely analytical approach, or to an imperfect matching of the fabricated couplers with the simulated design.

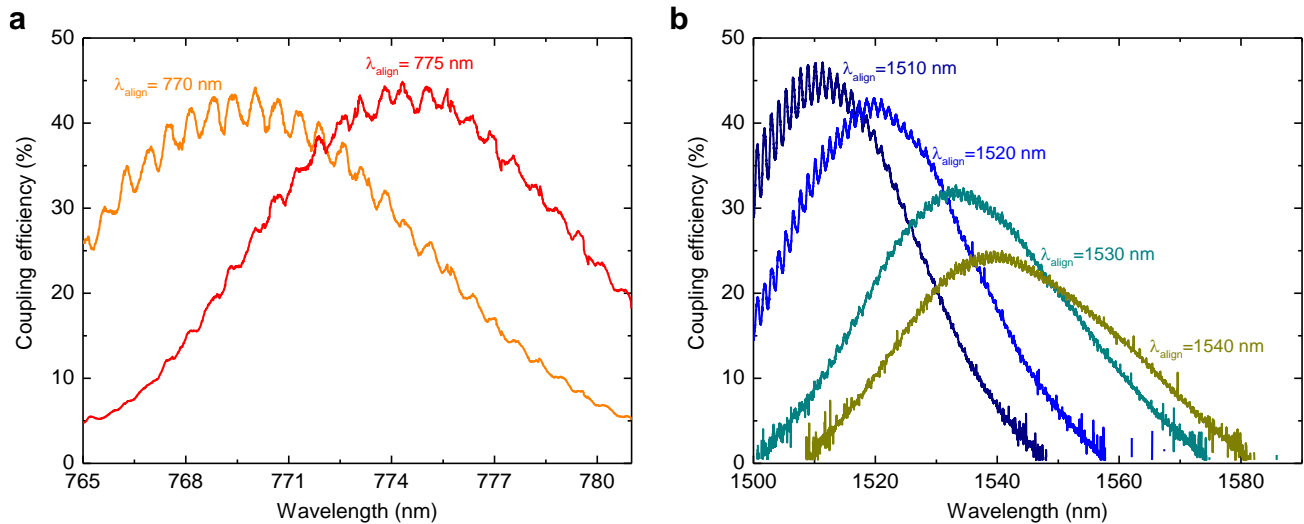

**Fig. 4. Transmission spectra of the fabricated optical device. (a)** For the same grating coupler operating in the near-visible wavelength range, we show two different spectra recorded at two different alignment wavelengths, here denoted as $\lambda_{align}$. Similar coupling efficiencies are measured. **(b)** For the same grating coupler operating in the infrared wavelength range, here we show four different spectra acquired at four different alignment wavelengths. A strong dependence of the coupling efficiency on the alignment wavelength $\lambda_{align}$ and a more dominant contribution of the back-reflected light at short wavelength correctly reflects the simulated behavior.

## CONCLUSION

We have presented efficient concentric apodized grating couplers, characterized by a negative diffraction angle, which allows to focus the upward-emitted light at a distance of approximately 100-150 μm above the photonic chip. This imaging effect transfers into a coupling efficiency of the $TE_0$ mode as high as -3.27 dB and -3.48 dB at 1550 nm and 775 nm, respectively, very close to the simulated value.

We anticipate that our approach can facilitate the development of wafer-scale LN photonic circuits for four main reasons: *i)* the simplicity of the employed fabrication process results in grating couplers that show high-fabrication yield, and highly reproducible performance across the entire chip; *ii)* the possibility of easily adapting the GC geometrical parameters to match any LN film thickness and desired etching depth; *iii)* their compact size and out-of-plane coupling operation make the coupling position extremely flexible on the LN surface; *iv)* the long working distance of our GCs releases from the constraint of placing the fiber as close as possible to chip surface to achieve the maximal CE, condition that might be not ideal in certain applications.

Finally, FDTD simulations predict that, by complementing our gratings with a metal back-reflector underneath the $SiO_2$ layer for improved directivity, coupling efficiencies exceeding 95% can be achieved. Thus, our approach can also assist the development of hybrid integrated photonic circuits for quantum photonic technologies[15], by providing a powerful method for the realization of low-loss fiber-to-chip interfaces with LNOI waveguides.

**ACKNOWLEDGEMENTS**

F.L. acknowledges the support of the Humboldt Research Fellowship for Postdoctoral Researchers. This work was further funded by the Deutsche Forschungsgemeinschaft (DFG, German Research Foundation) through SFB 1459. We gratefully acknowledge support by the European Research Council through grant 724707. We further acknowledge funding for this work from the European Union's Horizon 2020 Research and Innovation Program (Fun-COMP project, #780848 and PHOENICS, #101017237).